\documentclass{isd}

\usepackage[utf8]{inputenc}
\usepackage[T1]{fontenc}
\usepackage{graphicx}
\usepackage{booktabs}
\usepackage{multirow}
\usepackage{array}
\usepackage{amsmath,amssymb,amsfonts}
\usepackage{etoolbox}
\usepackage{float}
\usepackage{adjustbox}
\usepackage{xcolor}

\appto{\bibsetup}{\setlength{\itemsep}{0pt}}

\begin{document}

\begingroup
\footnotesize

\title[Metadata Candidate Identification in Property Graph Schemas]
{From Embedded Properties to Trait Nodes: A Design Method for Identifying Reusable Metadata in Property Graph Schemas}

\author[Sa'd et al.]
{Yahya Sa'd, Renzo Angles, Vojt{\v e}ch Merunka, Roberto Garc{\'i}a, Karel Kl{\'i}ma, Pavel Ber{\'a}nek}

\email{sady@pef.czu.cz,rangles@utalca.cl,merunka@pef.czu.cz}

\institution{
Czech University of Life Sciences Prague,
Universidad de Talca,
Czech University of Life Sciences Prague and Czech Technical University in Prague,
Universidad de Talca,
Czech University of Life Sciences Prague,
Czech University of Life Sciences Prague
}

\city{
Prague,
Talca,
Prague,
Talca,
Prague,
Prague
}

\country{
Czechia,
Chile,
Czechia,
Chile,
Czechia,
Czechia
}

\maketitle

\endgroup
\begin{center}
\small
\textit{Accepted for presentation at the 34th International Conference on Information Systems Development (ISD 2026), Prague, Czechia.}
\end{center}

\begin{abstract}
Property-graph schemas often contain descriptive properties that recur across heterogeneous
nodes and edges, yet schema designers lack a clear method for deciding whether such properties
should remain embedded or be treated as reusable metadata structures. This paper addresses this
design-stage problem within a 5GNF-oriented modeling perspective by proposing a method for
identifying metadata candidates based on five criteria: cross-element occurrence, conceptual
independence, lossless externalization, reuse potential, and governance relevance. The method
classifies properties into trait candidates, embedded properties, and borderline cases using a
rule-based decision workflow.

The approach is illustrated using a running example from a library domain and examined through an
illustrative validation involving participant-based classification tasks in two schema contexts.
The results show that recurrence alone is not a sufficient basis for externalization and that
metadata-candidate identification requires semantic interpretation beyond frequency. The main
contribution of the paper is methodological: it provides a more explicit and systematic basis for
deciding when descriptive properties should be modeled as reusable metadata in property-graph
schemas.

\end{abstract}

\section{Introduction}

Property graphs are widely used to represent complex and highly connected data in modern
information systems \cite{Angles2018PropertyGraphs,Angles2023PGSchema,ISO39075GQL2023,Robinson2015GraphDatabases}.
Their combination of graph structure and flexible properties makes them suitable for domains in
which entities, relationships, and contextual descriptions must be modeled together. However, as schemas
grow, some descriptive properties recur across different node and edge types. Although such properties are often stored as embedded attributes, some function more like reusable metadata than instance-specific data. Repetition of such descriptors can reduce schema clarity and
make shared descriptive structures harder to manage consistently
\cite{Gafurova2020MetadataNormalization,Koh2019MetadataNormalization,Ibrahim2023GraphDatabaseDesign}.

Recent studies have adapted classical normalization ideas to property graphs
\cite{Codd1970,Frisendal2022GNF,Merunka2024EOMAS,Skavantzos2023NormalizingPG,Skavantzos2025BCNF3NFPG}.
Within this line of work, Fifth Graph Normal Form (5GNF) extends normalization toward metadata
by allowing recurring and semantically independent descriptive structures to be externalized into
reusable Trait Nodes connected through explicit \texttt{HAS\_TRAIT} relationships
\cite{Sad2026ENASE5GNF}. This perspective improves schema organization but leaves open an earlier design question: how should a designer decide which descriptive properties should remain
embedded and which should be treated as metadata candidates?

This paper addresses that question by proposing a design method for identifying reusable metadata
candidates in property-graph schemas from a 5GNF-oriented perspective. The method uses explicit
criteria and a rule-based workflow to classify properties as trait candidates, embedded properties, or borderline cases. It is illustrated through a running example from a library domain and
examined through an illustrative validation across two schema contexts. The contribution is
methodological: a more explicit and systematic basis for deciding when descriptive properties
should be modeled as reusable metadata in property-graph schemas.

The remainder of the paper is organized as follows. Section~2 presents the background and
motivation. Section~3 introduces the running example. Section~4 presents the design method.
Section~5 applies the method to the example. Section~6 reports the evaluation. Section~7 reviews
related work. Section~8 concludes the paper.

\section{Background and Motivation}

\subsection{Property Graph Schemas and Reusable Descriptive Structures}

Property graphs combine graph structure with properties attached to nodes and edges, which makes
them effective for modeling interconnected data in many information-system settings
\cite{Angles2018PropertyGraphs,Angles2023PGSchema,ISO39075GQL2023,Bechberger2021GraphDatabasesInAction}.
During schema design, however, descriptive properties are often introduced locally and repeated
across different node and edge types, even when they express similar meaning.

Some of these repeated properties function less as ordinary instance-level attributes and more as
shared descriptive structures. They may carry contextual, classificatory, geographic, temporal,
provenance, or access-related meaning across several graph elements
\cite{Gafurova2020MetadataNormalization,Koh2019MetadataNormalization,Sadoughi2025MetaPropertyGraph}.
When such descriptions remain repeatedly embedded, the schema may become less modular and harder
to maintain.

\subsection{5GNF-Oriented Modeling Perspective}

Recent research has extended normalization ideas from the relational model to property graphs
\cite{Codd1970,Frisendal2022GNF,Merunka2024EOMAS,Skavantzos2023NormalizingPG,Skavantzos2025BCNF3NFPG,Sad2026ENASE5GNF}.
Within this line of work, 5GNF addresses reusable metadata by allowing recurring and semantically
independent descriptive structures to be externalized into reusable Trait Nodes connected by
explicit \texttt{HAS\_TRAIT} relationships \cite{Sad2026ENASE5GNF}.

In this view, traits are not ordinary domain entities. They represent how graph elements are
described rather than what those elements are. A descriptive property is therefore a plausible
metadata candidate when its meaning is not tied to a single graph element and can be externalized
without semantic loss.

\subsection{Research Gap and Paper Objective}

Although the 5GNF perspective provides a rationale for metadata normalization, it does not fully
solve the earlier design problem of deciding which properties are metadata candidates in the first
place. In practice, some cases are clear, but many depend on domain semantics and modeling intent.

This step matters because poor identification leads to poor design choices. Over-externalization
may fragment the schema, while under-externalization may leave reusable metadata hidden inside
ordinary properties. What is needed, therefore, is a practical and explainable method for
identifying trait candidates before normalization decisions are applied \cite{Peffers2007DSRM}.

The objective of this paper is to address that need by proposing a method for identifying reusable
metadata candidates in property-graph schemas and supporting design-stage decisions about when
trait-based externalization is semantically justified.

\section{Running Example}

\subsection{Baseline Schema}

\begin{figure*}[t]
    \centering
    \vspace{-4pt}
    \includegraphics[width=0.92\textwidth]{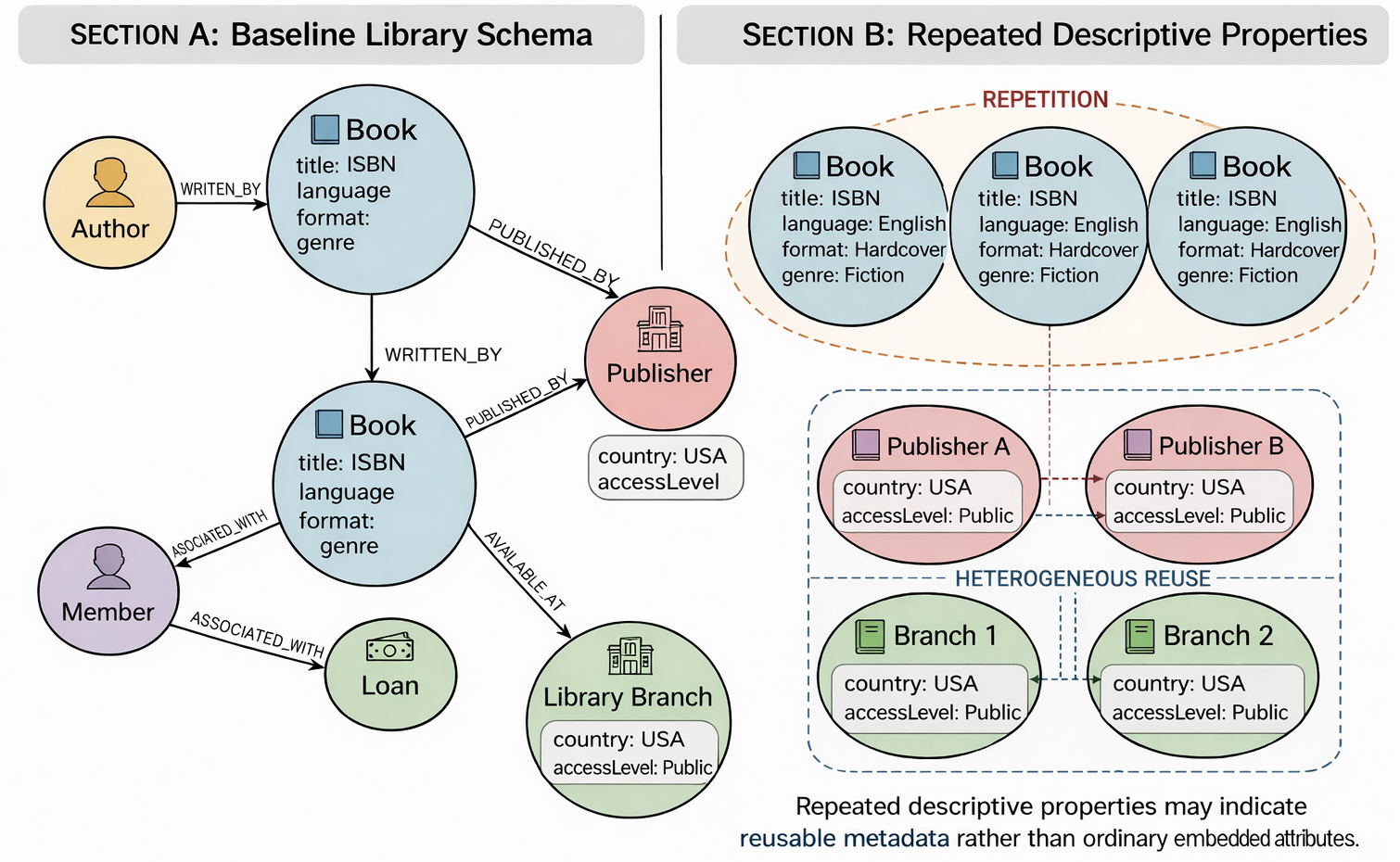}
    \caption{Baseline library schema with repeated descriptive properties.}
    \vspace{-4pt}
    \label{fig:baseline-library-schema}
\end{figure*}

To illustrate the problem, we use a compact running example in the library domain.
As shown in Fig.~\ref{fig:baseline-library-schema}, the baseline schema (on the left) contains node types such as \textit{Book}, \textit{Author}, \textit{Publisher},
\textit{LibraryBranch}, \textit{Member}, and \textit{Loan}, together with relationships such as
\textit{WRITTEN\_BY}, \textit{PUBLISHED\_BY}, and \textit{AVAILABLE\_AT}.
In addition, the figure shows (on the right) descriptive properties that recur across different graph
elements, including \textit{language}, \textit{country}, \textit{format}, \textit{genre}, and
\textit{accessLevel}.

The example is intentionally simple. Its purpose is not to model the full library domain, but to
provide a clear setting in which repeated descriptive properties can be examined as potential
metadata candidates.

\subsection{Metadata Candidate Ambiguity}

Several descriptive values in the library schema may recur across different elements. For example,
many \textit{Book} nodes may share the same \textit{language}, different \textit{Publisher} and
\textit{LibraryBranch} nodes may share the same \textit{country}, and properties such as
\textit{format}, \textit{accessLevel}, or \textit{preservationStatus} may also reappear in
multiple places.

However, repetition alone does not justify externalization. Some properties, such as
\textit{title}, \textit{ISBN}, and \textit{memberId}, remain closely tied to specific graph
elements and are better treated as embedded attributes. Others, such as \textit{language},
\textit{country}, and \textit{genre}, are more plausible candidates for trait-based
externalization. There are also borderline cases, such as \textit{publicationYear}, whose final
treatment depends on semantics and modeling intent.

This ambiguity motivates the design problem addressed in the paper: not only whether a value is
repeated, but whether it should remain an ordinary property or be treated as reusable metadata.

\section{Design Method for Metadata Candidate Identification}

\subsection{Method Overview}

The proposed method supports an early design decision in property-graph modeling: whether a
descriptive property should remain embedded or be treated as a reusable metadata candidate. Its
aim is not to automate this judgment fully, but to provide a structured basis for reasoning about
descriptive properties before normalization decisions are made.

The method takes as input a property-graph schema and the descriptive properties that appear
across its node and edge types. These properties are evaluated through five criteria: 

(C1) Cross-Element Occurrence;
(C2) Conceptual Independence;
(C3) Lossless Externalization;
(C4) Reuse Potential; and
(C5) Governance and Evolution Relevance.

The first three criteria form the semantic core of the method, since they define the minimum conditions under which trait-based externalization is semantically justified. The remaining two address the practical
design value of externalization through reuse and governance considerations.

The method produces a three-way classification: \emph{trait candidate}, \emph{embedded property},
or \emph{borderline case}. In practical use, the designer evaluates each property against
Criteria C1--C5 using the three-level scoring scale introduced below, then interprets the result
through the rule-based classification procedure. In this way, the method structures semantic
judgment into a repeatable sequence of design steps.

The criteria are illustrated below using the library-domain running example, especially through
properties such as \textit{language}, \textit{country}, \textit{title},
\textit{publicationYear}, and \textit{preservationStatus}, which together represent strong, weak,
and borderline support for trait-based externalization.

\subsection{Criterion C1: Cross-Element Occurrence}

The first criterion asks whether a descriptive property, or the descriptive concept expressed by
its values, appears across multiple graph elements rather than only in a single isolated context.
This matters because reusable metadata typically becomes visible when the same descriptive meaning
recurs in different parts of a schema.

In the running example, a property such as \textit{country} may appear in connection with
\textit{Publisher} and \textit{LibraryBranch}. Similarly, \textit{language} or
\textit{accessLevel} may recur across several node types. Such recurrence suggests that the
property may represent more than a purely local detail and may therefore deserve closer
examination as a metadata candidate.

At the same time, cross-element occurrence should be treated only as an initial signal. A
property may be repeated for practical reasons while still remaining semantically local to each
element in which it appears. Accordingly, this criterion helps identify plausible candidates, but
it does not by itself justify trait-based externalization. In this method, C1 is assessed
primarily at the schema level, while instance frequency may serve only as a contextual refinement
in borderline cases. A property may also recur across several node types while remaining
infrequent in practice if those node types have few instances. For the present method, recurrence
across node types is treated as the more informative case for trait-based externalization.

\subsection{Criterion C2: Conceptual Independence}

The second criterion asks whether a descriptive property refers to a concept whose meaning is not
fully dependent on a single graph element. A property exhibits conceptual independence when it can
be understood as expressing a descriptive concept in its own right, rather than merely recording
an internal detail of one specific node or edge.

This distinction is important because repetition alone is not enough. A property may appear many
times across a schema and still fail to represent reusable metadata if its meaning remains tied to
the individual element that carries it. By contrast, when a property refers to a concept that
retains the same descriptive role across multiple occurrences, it becomes a stronger candidate for
externalization.

In the running example, \textit{language} and \textit{country} are typical examples of properties
that may exhibit conceptual independence. Their meanings remain understandable across different
uses and are not confined to one particular element. By contrast, properties such as
\textit{title}, \textit{ISBN}, and \textit{memberId} are usually bound to the identity of a
specific element and therefore do not naturally function as reusable descriptive concepts.

For operational use, each criterion is evaluated on a three-level scale: \(1\) when the criterion
is clearly satisfied, \(0.5\) when it is only partially or context-dependently satisfied, and
\(0\) when it is not satisfied. This scoring does not replace semantic judgment, but it provides
a more explicit and repeatable basis for comparing properties across the schema.

In practical use, a score of \(1\) indicates that the criterion is satisfied in a clear and
stable way across the schema context under consideration. A score of \(0.5\) indicates partial,
mixed, or context-dependent support, that is, cases in which the criterion is plausible but not
strong enough to justify an unqualified positive judgment. A score of \(0\) indicates that the
criterion is clearly not satisfied. This scale is intended to reduce purely intuitive decisions by
making intermediate cases explicit rather than forcing a binary yes-or-no assessment.

\subsection{Criterion C3: Lossless Externalization}

The third criterion asks whether a property can be externalized into a reusable metadata
structure without losing its intended meaning or weakening the semantics of the original schema
element. This criterion is crucial because even a repeated and conceptually stable property may
still be a poor candidate for externalization if moving it out of its local context produces an
artificial or distorted representation.

A property satisfies this criterion when its descriptive role can be preserved after
externalization. In other words, the property should still mean the same thing when represented
through a reusable structure rather than as an embedded attribute.

In the running example, properties such as \textit{language}, \textit{country}, and
\textit{genre} can often be externalized without semantic loss, because their meanings remain
clear when represented as shared descriptors. By contrast, \textit{title}, \textit{ISBN}, and
\textit{memberId} would normally lose their intended role if separated from the elements to which
they belong. This criterion therefore acts as a safeguard: it ensures that externalization is
guided by semantic preservation rather than by recurrence alone.

\subsection{Criterion C4: Reuse Potential}

The fourth criterion considers whether externalization is likely to support meaningful reuse
within the schema. Whereas the first three criteria focus on semantic justification, this
criterion asks whether explicit reusable representation would also provide practical modeling
value.

Some properties are worth externalizing not only because they can be externalized without semantic
loss, but because doing so makes the schema cleaner, more modular, and easier to extend. In such
cases, externalization allows the same descriptive structure to be used in multiple parts of the
schema without repeated local definition.

In the running example, properties such as \textit{language}, \textit{country}, and
\textit{genre} may have clear reuse potential because the same descriptors can be associated with
many books and, depending on the design, with other elements as well. By contrast, some repeated
properties may offer little real benefit when externalized. This criterion therefore asks whether
reusable representation adds practical value rather than merely reducing duplication. A stronger
reuse signal arises when a property recurs not only within one schema but across multiple
application domains. The expected size of the resulting Trait Node population is also relevant,
since a property with only a very small number of distinct values may not justify the structural
overhead of externalization.

\subsection{Criterion C5: Governance and Evolution Relevance}

The fifth criterion asks whether a descriptive property is likely to matter for consistency,
controlled interpretation, or future schema evolution. Some properties become especially useful as
explicit metadata structures because their meanings need to remain stable across different parts of
the schema and over time.

This issue is relevant in growing information systems, where schemas are extended, new node types
are introduced, and descriptive categories are reused in new contexts. A property that seems local
in an early design may later require more systematic handling if it becomes important for shared
interpretation or maintenance.

In the running example, properties such as \textit{language}, \textit{country},
\textit{accessLevel}, and \textit{preservationStatus} may have this kind of relevance. Modeling
them only as local embedded attributes may make later maintenance more difficult if the same
descriptive meaning must be preserved consistently across the schema. This criterion does not mean
that every governed property must be externalized, but it highlights cases in which trait-based
representation may offer longer-term design benefits.

\subsection{Why These Five Criteria?}
Criteria C1--C3 correspond to the semantic admissibility conditions of
5GNF-oriented metadata externalization: cross-context recurrence (C1),
conceptual independence (C2), and lossless separation (C3)
\cite{Sad2026ENASE5GNF}. Criteria C4 and C5 capture design-level utility,
namely reuse and governance relevance. This separation preserves the
distinction between semantic validity and practical modeling benefit.

\subsection{Decision Workflow}
\begin{figure}[t]
    \centering
    \vspace{-4pt}
    \includegraphics[width=0.9\linewidth]{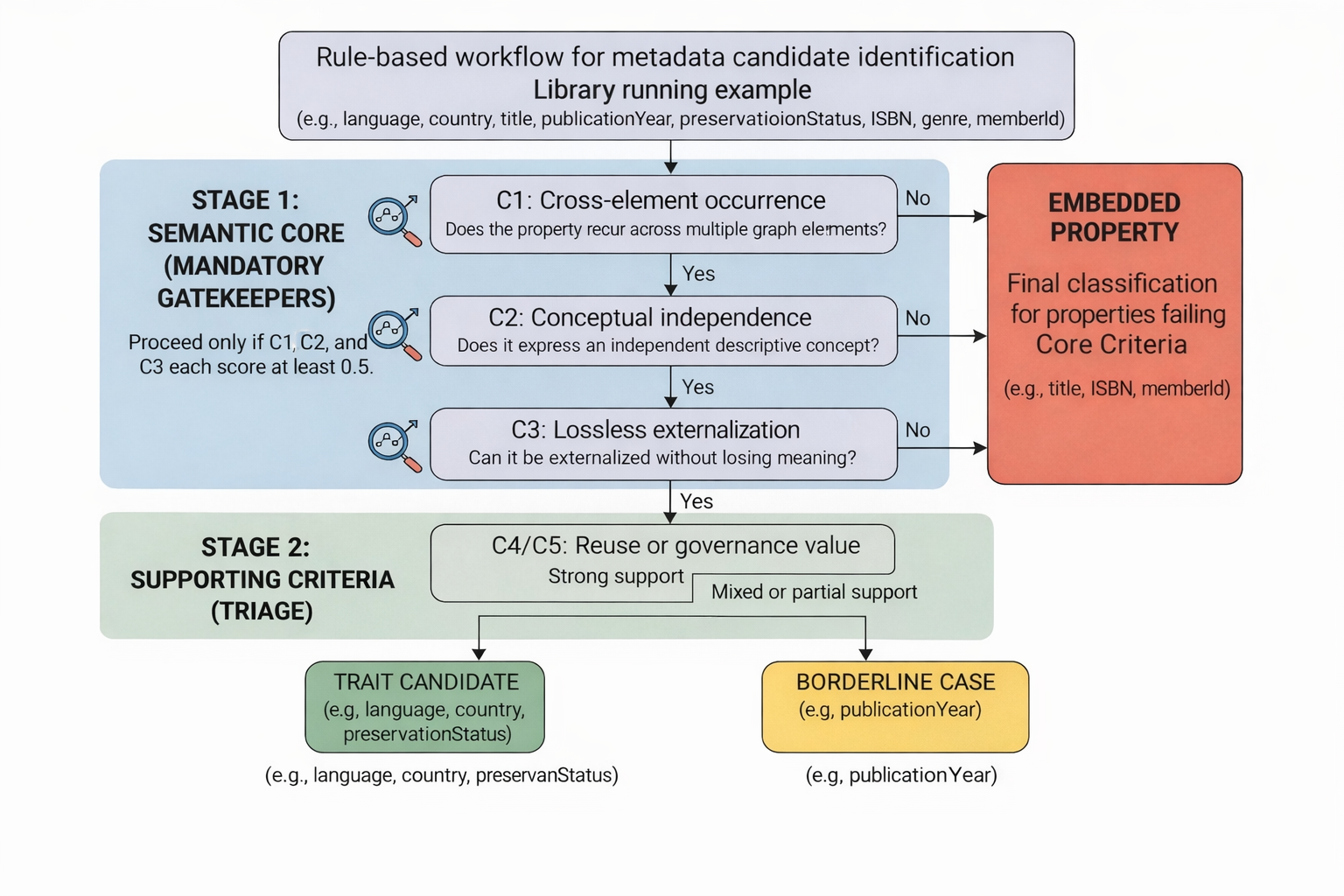}
    \caption{Rule-based workflow for metadata-candidate identification.}
    \vspace{-4pt}
    \label{fig:decision_workflow_library}
\end{figure}

The criteria are applied as a rule-based workflow rather than as independent
checks. Figure~\ref{fig:decision_workflow_library} summarizes the process
for the library running example.

The workflow first evaluates the semantic core (C1--C3). If this core is
not satisfied, the property remains embedded. If it is satisfied, C4 and C5
are used to determine whether externalization also provides sufficient
schema-level value. The method is therefore rule-based rather than
score-aggregative, since strong practical benefit should not override failure
of the semantic core.

\subsection{Classification Outcomes}
The method produces three outcomes: \emph{trait candidate},
\emph{embedded property}, and \emph{borderline case}.

For a property $p$, let $C_i(p) \in \{0,0.5,1\}$ denote the evaluation of
criterion $i$. A property is classified as a \emph{trait candidate} iff
\[
(C1(p)\geq 0.5 \land C2(p)\geq 0.5 \land C3(p)\geq 0.5)\land(C4(p)=1 \lor C5(p)=1).
\]
A property is classified as an \emph{embedded property} iff
\[
C1(p)=0 \land C2(p)=0 \land C3(p)=0.
\]
All remaining cases are treated as \emph{borderline cases}.

Operationally, the procedure is: (1) evaluate C1--C5; (2) apply the
decision rule above; and (3) refine borderline cases using contextual
factors such as value-domain size, reuse scope, optionality, uniqueness,
and governance needs. This preserves the distinction between semantic
admissibility and practical design value.

\section{Application to the Running Example}

\subsection{Property-by-Property Classification}

We now apply the proposed method to the library running example in order to classify selected
descriptive properties as trait candidates, embedded properties, or borderline cases. This
application is performed at the schema-design level and should be read as a worked example of the
method under the assumptions of the present schema.

Table~\ref{tab:running-classification} summarizes the classification results. The table is not
intended to cover every property in the schema, but to illustrate the main outcomes of the method.
Properties such as \textit{language}, \textit{country}, and \textit{genre} show strong support for
trait-based externalization. Properties such as \textit{title}, \textit{ISBN}, and
\textit{memberId} remain embedded because their meaning is tied to specific graph elements.
\textit{publicationYear} is treated as a borderline case, while \textit{preservationStatus} is
classified here as a trait candidate, although it remains more context-sensitive than the other
positive cases.

This application shows that repetition alone is not sufficient for externalization: a
frequency-based view could incorrectly treat repeated properties such as \textit{title} or
\textit{memberId} as metadata candidates, whereas the proposed method distinguishes repetition
from semantic reusability. The decision depends on whether a property expresses a reusable
descriptive concept and whether explicit representation provides semantic and design value.

\begin{table}[t]
\centering
\caption{Classification of selected descriptive properties in the running example}
\label{tab:running-classification}
\small
\setlength{\tabcolsep}{4pt}
\renewcommand{\arraystretch}{1.05}
\begin{adjustbox}{max width=\textwidth}
\begin{tabular}{|l|c|c|c|c|c|l|}
\hline
\textbf{Property} & \textbf{C1} & \textbf{C2} & \textbf{C3} & \textbf{C4} & \textbf{C5} & \textbf{Classification} \\
\hline
language & 1 & 1 & 1 & 1 & 1 & Trait candidate \\
\hline
country & 1 & 1 & 1 & 1 & 1 & Trait candidate \\
\hline
genre & 1 & 1 & 1 & 1 & 0.5 & Trait candidate \\
\hline
title & 0 & 0 & 0 & 0 & 0 & Embedded property \\
\hline
ISBN & 0 & 0 & 0 & 0 & 0 & Embedded property \\
\hline
memberId & 0 & 0 & 0 & 0 & 0 & Embedded property \\
\hline
publicationYear & 1 & 0.5 & 0.5 & 0.5 & 0.5 & Borderline case \\
\hline
preservationStatus & 1 & 0.5 & 0.5 & 1 & 1 & Trait candidate \\
\hline
\end{tabular}
\end{adjustbox}
\end{table}

The classification in Table~\ref{tab:running-classification} is based on the semantic role of each
property in the running example rather than on observed instance statistics. In practice, a
designer may refine the judgment further by considering expected data populations, governance
needs, value-domain size, and likely schema evolution.

\subsection{Resulting Design View}

The resulting design view contains three groups of descriptive properties. The first group includes
\textit{language}, \textit{country}, \textit{genre}, and \textit{preservationStatus}, which are
treated in this example as reusable metadata candidates. The second group includes
\textit{title}, \textit{ISBN}, and \textit{memberId}, which remain embedded because their meaning
is local to particular graph elements. The third group is represented by
\textit{publicationYear}, whose treatment depends more strongly on modeling intent and domain
interpretation.

The resulting design view therefore does not force externalization of all repeated descriptors.
Instead, it distinguishes properties that clearly support trait-based reuse from those that remain
local, while preserving a borderline category for context-dependent cases.

\section{Illustrative Validation}

\subsection{Validation Design}

The illustrative validation was designed to examine whether the proposed method can support
consistent and explainable reasoning when applied by different raters to repeated descriptive
properties in property-graph schemas. The purpose was not to provide strong empirical validation
of the method, but to assess whether the five-criterion rubric and the rule-based classification
procedure could be applied in a sufficiently comparable and interpretable way across a small set of
schema-design tasks.

The validation therefore focused on agreement patterns rather than on statistical generalization.
In particular, it examined whether raters converged more strongly on clearly reusable descriptive
properties and clearly local embedded properties, while showing greater variation on semantically
ambiguous cases. This focus is consistent with the purpose of the method, which is to support
structured design judgment rather than to replace it with a fully automatic decision rule.

For this reason, the study was framed as illustrative validation. It was intended to provide
initial evidence that the method can function as a shared semantic rubric for design reasoning,
while also revealing which kinds of properties remain more sensitive to domain interpretation in
practice.

\subsection{Materials and Tasks}

The validation used two compact schema contexts derived from the paper: the library-domain schema
used as the running example and a second research-information schema. These two cases were chosen
to include both relatively clear and more contestable descriptive properties, so that the method
could be examined across different semantic conditions.

From the two schemas, ten properties were selected for evaluation. The library-domain task
included \textit{language}, \textit{country}, \textit{title}, \textit{publicationYear}, and
\textit{preservationStatus}. The research-information task included \textit{fundingAgency},
\textit{publicationType}, \textit{title}, \textit{projectStatus}, and \textit{grantId}. Together,
these properties were intended to cover likely trait candidates, clearly embedded properties, and
borderline cases.

Each participant received the same packet. For each property, participants were asked to evaluate
Criteria C1--C5 using the response options Yes, Partial, and No, then assign one final class:
Trait candidate, Embedded property, or Borderline case. They were also asked to provide a short
written justification for each classification. This structure enabled comparison of both the final
decisions and the underlying reasoning.

To support transparency and reproducibility, the full experiment packet and anonymized response
materials will be made publicly available in an online repository upon acceptance of the paper.
\subsection{Participants}

The study involved 5 participants with backgrounds in databases, information systems, and
conceptual modeling. Their professional experience ranged from 4 to over 20 years. Although not
all participants specialized in property-graph databases, the task primarily required semantic
reasoning about schema design rather than advanced graph-specific implementation expertise.

This participant profile is consistent with the intended scope of the method. The proposed
approach is designed to support design-stage judgment about descriptive properties and reusable
metadata, and therefore depends mainly on general modeling competence rather than on narrow
technical specialization in a particular graph technology.

\subsection{Procedure}

All participants received the same experiment packet and completed the task independently. They
were first asked to read the plain-language description of the five criteria (C1--C5) and review
the two schema figures. They then evaluated each of the ten selected properties.

For each property, participants marked one response for each criterion (Yes, Partial, or No),
assigned a final classification (Trait candidate, Embedded property, or Borderline case), and
provided a short written justification. No discussion between participants was allowed during the
task.

After collection, the responses were prepared for analysis by mapping Yes, Partial, and No to
numerical scores (1, 0.5, and 0, respectively). The same rule-based decision procedure defined in
the method was then applied consistently across all responses to derive comparable classification
outcomes.

\subsection{Results}

Across the returned response sheets, the strongest convergence appeared for clearly local
identifier-like or instance-specific properties. In both schema contexts, properties such as
\textit{title}, \textit{ISBN}, \textit{memberId}, and \textit{grantId} were generally treated as
embedded properties, with participants repeatedly justifying this choice by emphasizing local
meaning, owner-specific interpretation, or lack of reuse value. This pattern suggests that the
method supports relatively stable judgments in cases where descriptive meaning remains closely tied
to a specific graph element.

Clear positive cases also emerged, although with less uniformity than the embedded cases.
Properties such as \textit{fundingAgency} were commonly treated as good candidates for
externalization because they were seen as reusable descriptive categories with independent
meaning. In the library case, \textit{country} was also often interpreted positively, but some
participants treated it more cautiously and linked the decision to expected queries, reuse
frequency, or broader domain assumptions. This indicates that the method can also support
convergence on plausible trait candidates, while still leaving room for justified contextual
variation.

The greatest variation appeared in semantically contestable cases. In particular,
\textit{publicationYear}, \textit{preservationStatus}, \textit{publicationType}, and
\textit{projectStatus} produced more divergent judgments. Some participants treated these
properties as plausible trait candidates when future governance, controlled vocabularies, or
schema extension were taken into account, whereas others classified them as borderline or kept
them closer to local interpretation because they appeared too context-dependent, too small in
value domain, or insufficiently reusable in the schema as shown. This concentration of variation
in borderline cases suggests that disagreement was not random, but arose primarily where semantic
interpretation and modeling intent played a stronger role.

The written justifications also revealed a recurring interpretive distinction. Some participants
evaluated properties mainly from the visible schema structure, especially whether they appeared
across multiple node types. Others reasoned more broadly from likely data populations,
anticipated queries, governance value, or possible future development of the domain. This
difference helps explain why agreement was highest for clearly local properties and lower for
context-sensitive descriptors, while also indicating that the method provided a shared decision
structure even when raters emphasized different contextual considerations.

\subsection{Interpretation}

The results support the main claim of the paper: metadata-candidate 
identification cannot be reduced to repetition alone and instead 
requires structured semantic judgment. Participants showed high 
convergence on clearly local properties, while variation increased 
for properties whose interpretation depends on reuse expectations, 
governance needs, or domain context.

The responses also reveal two complementary interpretive perspectives. 
Some participants evaluated properties primarily from the visible schema 
structure; others reasoned from expected data populations, anticipated 
queries, and likely schema evolution. This distinction explains the 
divergence on properties such as \textit{language}, \textit{country}, 
and \textit{preservationStatus}.

Participant comments further clarified the boundaries of the method. 
Cross-element occurrence (C1) was interpreted differently depending on 
whether it was assessed at the schema level or the instance level, which 
suggests that C1 captures structural recurrence whereas instance 
frequency may act as an additional contextual factor. Participants also 
noted that value-domain size, optionality, and uniqueness can influence 
borderline judgments, and that recurrence across node types was 
generally considered more relevant than recurrence across edge types.

Taken together, these observations reinforce the role of the method as 
a structured design aid. The five criteria provide a schema-level 
semantic framework, while factors such as instance distribution and 
schema constraints act as contextual refinements for borderline cases.

\subsection{Limitations}

The validation presented in this paper is intentionally limited in scope. The small number of
participants reflects the illustrative nature of the study. The goal was not statistical
generalization, but to examine whether the proposed criteria can support comparable and explainable
reasoning across different individuals. The observed patterns of agreement and disagreement
therefore provide insight into how the method behaves in both clear and ambiguous cases.

A second limitation concerns the level of analysis. The method operates at the schema-design level
and does not incorporate instance-level statistics, such as data distributions or frequency of
property values. As highlighted by participant feedback, such factors may influence how strongly a
property is perceived as reusable in practice, especially in borderline cases.

Finally, the classification of some properties remains context-dependent. Properties such as
\textit{publicationYear} or \textit{preservationStatus} may be interpreted differently depending on
domain assumptions, expected queries, or governance requirements. This variability reflects the
inherent role of semantic judgment in schema design rather than a deficiency of the method itself.

\section{Related Work}

\subsection{Graph Normalization}

The proposed method is grounded in the broader tradition of normalization, which originated in
the relational model as a response to redundancy and dependency-related anomalies
\cite{Codd1970}. Later work extended normalization to graph-based settings, including graph
normal forms and conceptual normalization for graph databases
\cite{Frisendal2022GNF,Merunka2024EOMAS}. More recent research has shown that dependency-based
normalization can also be formulated for property graphs, including forms analogous to Third
Normal Form and Boyce--Codd Normal Form
\cite{Skavantzos2023NormalizingPG,Skavantzos2025BCNF3NFPG}.

These works provide the normalization background of the present paper, but they mainly address
how graph schemas can be normalized once relevant structures have already been identified. Recent
trait-based work extends this trajectory to metadata through 5GNF, which externalizes recurring
metadata into reusable Trait Nodes \cite{Sad2026ENASE5GNF}. By contrast, the present paper focuses
on the earlier question of identifying which repeated descriptive properties should enter the scope
of such treatment.

\subsection{Property Graph Schema Design}

Property-graph schema design has received increasing attention as graph databases have matured
from flexible implementation platforms into systems that also require explicit schema reasoning
\cite{Angles2018PropertyGraphs,Angles2023PGSchema,ISO39075GQL2023}. Existing work has clarified
the role of schema constraints, type structures, and modeling support for property-graph systems.
However, it does not directly address how repeated descriptive properties should be distinguished
from reusable metadata structures during design.

\subsection{Metadata and Semantic Modeling}

The problem addressed in this paper is also related to prior work on metadata representation,
normalization, and semantic modeling. Research on metadata-aware graph representations and
reification has shown that descriptive information may require explicit structures beyond ordinary
local properties \cite{Hartig2017RDFStar,Koh2019MetadataNormalization,Sadoughi2025MetaPropertyGraph}.
Other work has emphasized the organizational value of metadata normalization in digital-library
settings \cite{Gafurova2020MetadataNormalization}.

Related work from ontology and metadata-standardization communities is likewise relevant.
Ontology-based approaches provide explicit semantic structures, while metadata-registry frameworks
support standardized definition and governance of metadata elements across domains
\cite{PROVO2013,DCMITerms,ISO11179_1_2023}. These lines of work show how metadata can be
represented, standardized, and governed once recognized, but they do not provide a property-graph
design method for identifying which repeated descriptive properties should be treated as reusable
metadata candidates before normalization.

Table~\ref{tab:method-comparison} summarizes the position of the proposed method relative to these
approaches.

\begin{table}[t]
\centering
\caption{Position of the proposed method relative to related approaches}
\label{tab:method-comparison}
\scriptsize
\setlength{\tabcolsep}{3pt}
\renewcommand{\arraystretch}{1.04}
\begin{adjustbox}{max width=\columnwidth}
\begin{tabular}{|p{1.9cm}|p{2.0cm}|p{1.4cm}|p{1.7cm}|p{3.7cm}|}
\hline
\textbf{Approach} & \textbf{Main focus} & \textbf{Stage} & \textbf{Output} & \textbf{Difference from this paper} \\
\hline
Relational normalization &
Redundancy removal through normal forms and decomposition &
Logical schema design &
Higher normal forms &
Provides the normalization foundation, but not metadata-candidate identification in property-graph schemas. \\
\hline
Graph / conceptual graph normalization &
Normalization reasoning for graph databases and conceptual graph structures &
Graph or conceptual design &
Cleaner graph structures &
Extends normalization to graphs, but does not decide which repeated descriptive properties should become reusable metadata. \\
\hline
Property-graph normalization &
Dependency-aware normalization of property graphs &
Schema normalization &
Decomposed schemas &
Focuses on normalization outcomes rather than on the earlier identification of metadata candidates. \\
\hline
Metadata-aware graph representation / reification &
Explicit representation of metadata and annotations &
Representation modeling &
Richer metadata structures &
Shows how metadata can be represented explicitly, but not when repeated descriptive properties justify such treatment. \\
\hline
Ontology / metadata-registry approaches &
Standardized semantic representation and metadata governance &
Representation and metadata management &
Shared vocabularies and metadata definitions &
Supports metadata standardization and governance once identified, but not design-stage candidate identification in property-graph schemas. \\
\hline
\textbf{Proposed method} &
\textbf{Identification of reusable metadata candidates for 5GNF-oriented trait externalization} &
\textbf{Pre-normalization design stage} &
\textbf{Trait / embedded / borderline classification} &
\textbf{Addresses the earlier design question of whether a repeated descriptive property should remain embedded or enter the scope of trait-based externalization.} \\
\hline
\end{tabular}
\end{adjustbox}
\end{table}

\section{Conclusion}

This paper addressed the problem of identifying reusable metadata candidates in property-graph
schemas at the design stage. Instead of focusing on normalization as a transformation result, the
proposed method supports the earlier decision of whether descriptive properties should remain
embedded or be externalized as trait-based structures. The method introduces five criteria and a
rule-based workflow that together provide a structured basis for this decision.

The application of the method to the running example and its examination through an illustrative
validation show that repeated occurrence alone is not sufficient to justify externalization.
Properties must also demonstrate conceptual independence, support lossless externalization, and
provide meaningful reuse or governance value. At the same time, the results confirm that some
cases remain context-dependent, and that semantic interpretation remains an integral part of
schema design.

The main contribution of the paper is therefore methodological: it makes the process of
metadata-candidate identification more explicit, more systematic, and more transparent. By
clarifying the role of semantic criteria in this decision, the method provides a foundation for
more consistent schema design and opens the way for future tool support and broader empirical
evaluation in property-graph modeling.

\section*{Acknowledgment}
The author used an AI-based language assistant to improve clarity and structure. 
All ideas, methodological decisions, and final content were developed and verified by the authors.

\noindent
\textbf{Affiliations.}
Yahya Sa'D, Vojtěch Merunka, Karel Klíma, and Pavel Beránek are with Czech University of Life Sciences Prague, Czechia.
Renzo Angles is with Universidad de Talca, Chile.
Roberto García is with Czech Technical University in Prague, Czechia.

\bibliographystyle{plain}
\bibliography{references}

\end{document}